\documentclass[10pt]{article}
\usepackage{OSAmeet}

\newcommand{\beq}{\begin{equation}}
\newcommand{\beqa}{\begin{eqnarray}} 
\newcommand{\eeqa}{\end{eqnarray}} 
\newcommand{\eeq}{\end{equation}}
\newcommand{\dg}{^{\dagger}}
\def\reals{\hbox{\rm I\kern -.2em R}}
\def\sdd{Schr$\ddot{\mbox{o}}$dinger }
\def\sdds{Schr$\ddot{\mbox{o}}$dinger's }
\def\com{\hbox{\rm l\kern -.4em C}}
\def\tr{\mbox{Tr}}

\def\dfn{\hbox{\rm $=$\kern -.95em $~^{^{\triangle}}$}}
 
\begin{document}

\title{Entanglement in particle-detector interactions}
\begin{center}
Michael Steiner and R. W. Rendell \\
Naval Research Laboratory \\
202-404-1886, Michael.Steiner@nrl.navy.mil \\
\end{center}

\HeaderAuthorTitleMtg{Steiner et.al.}{Meas.QD}{ICQI 2001}

\begin{abstract}
It is predicted by \sdds equation that entanglement will occur in the interaction between detector and particle. We provide an analysis of the entanglement using the Gurvitz model of double-dot and detector.  New results on entangled doubled-dots are provided as well as implications on Quantum Information processing.  
\end{abstract}


\noindent                      

The prediction of entanglement under \sdd evolution is apparent when one considers the interaction of a particle with a detector.  One can see this by the well-known and simple argument that a particle will become correlated with the eigenstates of a measuring device. Hence if the particle were in an initial superposition of these eigenstates, then by invoking linearity, the result follows. A more general theoretical proof of the resulting entanglement was provided recently by Bassi and Ghirardi \cite{Q1:Bassi}. In this paper, we proceed further on investigating the dynamics of the predicted entanglement.

In order to answer this question a more realistic model of a detector is needed.  Gurvitz \cite{Q1:Gurvitz1} recently has made an important contribution in the study of the measurement problem in his analysis of an electron in a double-dot (DD) interacting with a quantum point contact (QPC).  His analysis considers the two components of the system: the current through the QPC and the density matrix of the electron. In this paper, we first provide a summary of results found in \cite{Q1:Steiner2} in which we extended Gurvitz's analysis by quantifying the dynamical entanglement that results between a single DD and QPC. We also provide new results in the consideration of two entangled DDs whereby one of the DDs interacts with the QPC. 

\section{Background}
\label{secback}

In order to examine the role of entanglement for this paper we utilize as a measure the entropy of entanglement of a pure composite bipartite state.  The entropy of entanglement $({\cal S}) $ of a bipartite pure state $\rho\in K(H_{A} \otimes H_{B})$ is given by Von Neumann's entropy of either $\rho_{A}$ or $\rho_{B}$, $ {\cal S}(\rho)  =   -\tr (\rho_{A} \log \rho_{A} )$

\begin{figure}
\centerline{\scalebox{.4}{\includegraphics{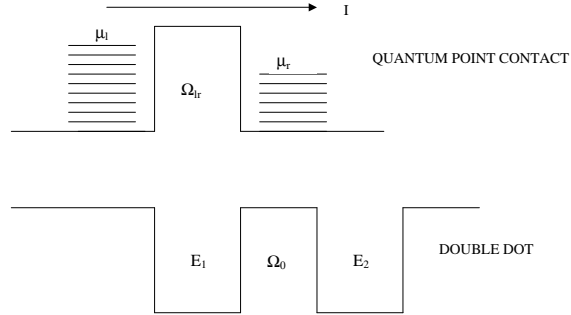}}}
\caption{Double Dot and Quantum Point Contact}
\label{figgur}
\end{figure}

Gurvitz \cite{Q1:Gurvitz1} considered the measurement of a single electron oscillating in a double-dot by using a quantum point contact detector.  We briefly review the setup in Fig. \ref{figgur}, but refer the reader to \cite{Q1:Gurvitz1}. The barrier shown in the figure is connected with two reservoirs at the potentials $\mu_{L}$ and $\mu_{R}=\mu_{L}-V_{d}$ respectively where $V_{d}$ is the applied voltage.  The Hamiltonian $H$ consisting of the point contact Hamiltonian $H_{PC}$, double-dot $H_{DD}$ and their interaction $H_{int}$ is
\[ H= H_{QPC} + H_{DD} + H_{int} \] where 
\begin{eqnarray*}
H_{QPC} & = & \sum_{l} E_{l} a_{l}\dg a_{l} +\sum_{r} E_{r} a_{r}\dg a_{r} + \sum_{l,r} \Omega_{lr}(a_{l}\dg a_{r}+ a_{r}\dg a_{l})  \\
H_{DD} & = & E_{1} c_{1}\dg c_{1} + E_{2} c_{2}\dg c_{2} + \Omega_{0}(c_{2}\dg c_{1} + c_{1}\dg c_{2}), \\
H_{int} & = & -\sum_{l,r} \Omega_{lr}c_{2}\dg c_{2}(a_{l}\dg a_{r}+ a_{r}\dg a_{l}), 
\end{eqnarray*}
and $E_{l,r}$ are the energy levels in the respective QPC reservoirs, $E_{1},E_{2}$ are the energy levels of the DD, $\Omega_{lr}$ is the coupling between the reservoirs, $\Omega_{0}$ is the coupling between the left and right dots, $a_{l}\dg$,$a_{l}$ denotes the respective creation and annhilation operators in the left reservoir (similarly for the right) and similarly $c_{i}\dg$, $c_{i}$ are creation and annihilation operators for the double-dot in state $i$.  A coupling parameter is defined as $\Gamma_{d}= T_{1} V_{d}/(2 \pi)$ where $T_{1}$ is the transmission coefficient between the DD and QPC.

\section{Entanglement for the DD-QPC}
\label{secent}

Suppose that the initial state of the DD is a pure state parameterized by $\theta,\phi\in[0,2\pi]$ such that $\sigma_{11}(0)= \cos^{2} \frac{\theta}{2}$, and $\sigma_{12}(0)= \sin \frac{\theta}{2} \cos \frac{\theta}{2} \exp(-i\phi)$. Then the entropy of entanglement, ${\cal S}$ is characterized by the eigenvalues of the density matrix $\sigma(t)$ of the DD. 

We define $\alpha\dfn \Gamma_{d}/\Omega_{0}$ and $\tau\dfn \Omega_{0} t$. Hence, $\alpha$ is a normalized measure of the coupling between the DD and QPC, and $\tau$ is normalized to $\Omega_{0}$.  The case where $\alpha>>1$ is strong coupling whereas for $\alpha<<1$ is weak coupling.  Entanglement between the DD and QPC and the rate of the entanglement, defined as ${\cal R}(\tau)\dfn \frac{d {\cal S}}{d\tau}$ was determined for several cases and for various couplings. Several results in \cite{Q1:Steiner2} are summarized below:

\begin{itemize}
\item Double-Dot initially in left Dot coupled to QPC, ($\theta=0$).
\begin{enumerate}
\item ${\cal S}(0)=0$. ${\cal S}(\tau)\rightarrow 0$ for all $\tau$ as $\alpha\rightarrow 0$.
\item There exists a nearly optimal coupling of $\alpha \approx 5$ that minimizes the entanglement time to approximately $\tau=1$. ${\cal S}(\tau)$ for a given $\tau$ decreases with coupling in the strong coupling regime
\item ${\cal R}(\tau)=0$ at $\tau=0$. ${\cal R}(\tau)\rightarrow 0$ as $\tau\rightarrow \infty$ and also decreases with increasing coupling for large $\tau$ 
\end{enumerate}
\item Double-Dot initially in a Superposition ($\theta=90$)
\begin{enumerate}
\item ${\cal S}(\tau)$ increases at a given $\tau$ with increased coupling. Entanglement time surpasses the optimal minimal entanglement time for the case of localised DD
\item The rate of entanglement ${\cal R}(\tau)>0$ at $\tau=0$ 
\end{enumerate}
\end{itemize}
Hence an initial superposition leads to an unbounded entanglement rate increase, whereas when the DD is initially localized, the rate is bounded due to the existence of an optimal coupling.

\section{DD-DD interacting with QPC}

\begin{figure}
\centerline{\scalebox{.4}{\includegraphics{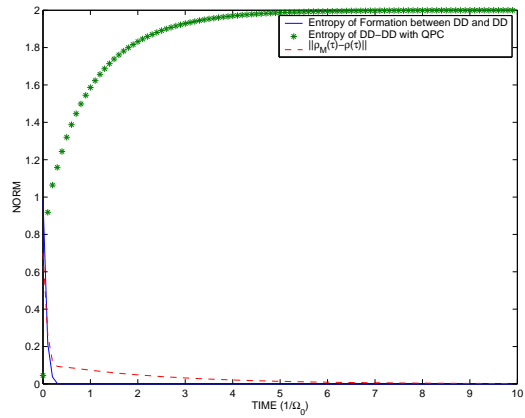}}}
\caption{Effect of QPC on DD-DD (Singlet, $\Gamma_{d} = \alpha \Omega_{0},\alpha=20$)}
\label{figb}
\end{figure}

We now consider double-dots that are  entangled {\it a priori} in a singlet state. One of the double dots, denoted by $DD_{1}$, interacts with the QPC.  Note that the Hamiltonian shown in Sec. 1 does not incorporate the effect of the entanglement with the second DD. The entanglement effects were computed by first characterizing the QPC-DD interaction in terms of a positive operator value expansion.  A technique similar to that developed in \cite{Q1:Chuang} was adapted for this purpose. There are several entanglement quantities that are of interest. In Fig. \ref{figb} we plot the \sdd predicted entanglement between the two DDs.  Although the initial state is pure, the state shortly becomes mixed and hence the measure used to quantify this entanglement was the Entropy of Formation, for which there is a closed form solution. Note that the two DDs disentangle under \sdd evolution in a very fast, microscopic, time period.  Although the two DDs are predicted to disentangle, it turns out that \sdds equation predicts that the two DDs, considered as a system, are predicted to entangle with the QPC as shown. Note that the entanglement time is substantially longer than the disentanglement time of the two DDs. 

In Ref. \cite{Q1:Steiner2} we left open the question of whether or not there exist conditions for which projective measurement occurs.  Note that if the QPC was a measuring device, then the measurement postulate should be applied, and there would be no entanglement between the two double-dots.  Hence, we analyze this problem and report on the results both in two ways: 1) treating the detector and entangled particles entirely according to \sdds equation and 2) assuming the QPC is a measurement device and applying the projection postulate.  

For the case of projective measurement where the result of the measurement is not known, the average density matrix $\rho_{M}$ is initialized at $t=0$ as follows:
\beq \label{eqnm}
\rho_{M}(0) = p_{L} \rho(DD_{1}=D_{L}) + p_{R} \rho(DD_{1}=D_{L}), \eeq
where $\rho(DD_{1}=D_{L})=(L \otimes \mbox{I}_{2}) \rho (L \otimes \mbox{I}_{2})'$ (re-normalized to unit trace as needed) and $L$ denotes the projection operator that projects to the left dot and $D_{L}$ is the resultant density operator of the DD upon projection, $\rho$ denotes the singlet density operator of the two DDs.  The probabilities are computed by $p_{L}=\tr((L \otimes \mbox{I}_{2}) \rho)$ and similarly for the right dot.  Now, if one initializes the DD-DD with the singlet state, and allows it to evolve, the result will be $\rho(t)$. On the other hand if the system is initialized on average as in (\ref{eqnm}) then the system will evolve, on average, as $\rho_{M}(t)$. We plot the difference between the average density matrix under the measurement postulate and $\rho$ as $||\rho_{M}(\tau)-\rho(\tau)||$ where $||\cdot ||$ denotes the Frobenius norm used for convenience, although other measures that relate to quantum distinguishability would be preferred. Note that the biggest difference is at $\tau=0$. However, asymptotically the two density operators converge. Note that actual detectors have a physical response time that also would have to be included in the computation of $\rho_{M}$.

Surprisingly, the DD-DD becomes disentangled both under \sdd evolution and via the measurement postulate. The conclusion that measurement and \sdds equation provide similar predictions was also found for a different experiment in Ref. \cite{Q1:Frerichs}. However, our result has shown a difference between measurement and \sdd evolution, albeit small. More work is needed to define experiments that better differentiate these cases.

{\bf \noindent Acknowedgment:}
The authors would like to thank Dr. P. Reynolds for ONR funding.

\end{document}